\documentclass[sigconf,screen,nonacm]{acmart}

\setcopyright{none}

\begin{document}

\title{Syndrome-based Calibration Using Gate-Error Amplification}

\author{Andrew T. Kamen}
\authornote{andrewkamen@uchicago.edu}
\affiliation{
  \department{Pritzker School of Molecular Engineering}
  \institution{University of Chicago}
  \city{Chicago}
  \state{IL}
  \country{USA}
}

\author{Arielle Sanford}
\affiliation{
  \department{Department of Computer Science}
  \institution{University of Chicago}
  \city{Chicago}
  \state{IL}
  \country{USA}
}

\author{Frederic T. Chong}
\affiliation{
  \department{Department of Computer Science}
  \institution{University of Chicago}
  \city{Chicago}
  \state{IL}
  \country{USA}
}

\author{Andy J. Goldschmidt}
\affiliation{
  \institution{Johns Hopkins Applied Physics Laboratory}
  \city{Laurel}
  \state{MD}
  \country{USA}
}

\author{Siddharth Dangwal}
\affiliation{
  \department{Department of Computer Science}
  \institution{University of Chicago}
  \city{Chicago}
  \state{IL}
  \country{USA}
}

\pagestyle{plain}

\begin{abstract}
We present Gate-Error Amplification for Recalibration with Syndromes (GEARS), a runtime calibration framework that leverages gate-error amplification to efficiently tune control parameters during quantum error correction. Maintaining reliable control of large qubit arrays throughout hours-long computations requires repeated calibration to compensate for hardware drift. In conventional calibration, targeted qubits are temporarily removed from the code, introducing non-negligible space-time overhead. An elegant alternative uses syndrome information already collected during error correction as feedback for in situ recalibration, but acquiring signal requires sufficient (1) measurement budget or (2) probe strength, i.e., the size of deliberate perturbations applied to the control parameters. More measurements improve statistical precision but result in slower control updates, while stronger perturbations produce larger calibration signals at the cost of additional errors.
GEARS augments syndrome-based calibration with an established gate characterization technique called gate-error amplification. Our approach repeats selected CZ gates and probes their controls in spatially separated groups, strengthening local calibration signals while limiting the logical cost of probing. Surface code simulations show that GEARS with depth-three amplification reduces measurement requirements by approximately a factor of eight at matched gradient SNR and fixed probe strength. In closed-loop calibration, GEARS with depth-three amplification reduces excess expected logical failures by approximately 31-40\% relative to calibration without amplification across the evaluated execution windows. Under time-dependent drift, GEARS increases the estimated critical drift frequency by approximately 1.9$\times$, extending the range of drift frequencies over which calibration reduces expected logical failures. These results indicate that deliberately amplifying gate errors during operation can improve the long-time performance of quantum computations.
\end{abstract}

\maketitle

\section{Introduction}\label{sec:introduction}

\begin{figure}[t]
    \centering
    \includegraphics[width=0.80\columnwidth]{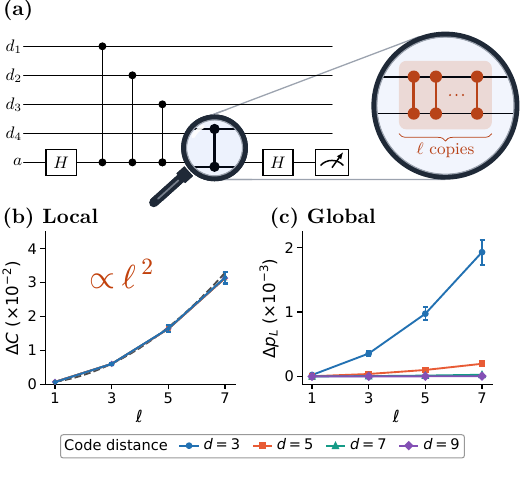}
    \caption{GEARS amplifies gate errors for efficient calibration during quantum error correction. (a) The selected CZ gate is applied $\ell$ times during a stabilizer measurement. (b) The mean increase in local detection rate, $\Delta C$, is approximately proportional to $\ell^2$ (dashed line) and is nearly independent of code distance, $d$. (c) $\Delta p_L$ is the increase in logical failure probability per round relative to calibrated operation without amplification. It decreases approximately exponentially with $d$. Points show means over 16 random directions of a 0.05-radian control offset on one CZ gate in a $Z$ check (see Sec.~\ref{subsec:cz-error-model} for the error model). Error bars show $\pm1$ standard error.}
    \label{fig:amplification-mechanism}
\end{figure}

\emph{Quantum Error Correction} (QEC)~\cite{gottesman1997stabilizer, shor_ftqc, Fowler_2012} is necessary to implement flagship quantum algorithms like quantum chemistry, factoring, and quantum dynamics simulations~\cite{shor1994algorithms, lloyd1996universal, aspuru2005simulated}. Resource estimates suggest that these computations could take a long time to complete~\cite{beverland2022assessing}. For example, Ref.~\cite{beverland2022assessing} estimates that quantum chemistry circuits could take weeks to centuries depending on the underlying quantum technology and the exact workload being executed. To ensure these programs execute successfully, QEC is used to maintain a low \emph{Logical Error Rate} (LER) throughout the computation.

QEC encodes logical qubit information in multiple physical qubits: data qubits collectively store the logical state, and ancilla qubits interact with nearby data qubits to measure their parities without directly measuring the encoded logical information (Fig.~\ref{fig:amplification-mechanism}a) \cite{Fowler_2012}. A \textit{detector} is a set of parity measurements whose combined parity is fixed in the absence of errors~\cite{sivak2026reinforcement, gidney2021stim}. Each time a detector detects a potential error, it outputs the bit ``1'' (a process called \emph{detector firing}); otherwise, it outputs ``0''. Detectors are sensitive to physical error rates and respond locally to errors (Fig.~\ref{fig:amplification-mechanism}b). The collective bit string generated by all detectors represents a global error signature or \emph{syndrome}. A \emph{decoder} is an algorithm that infers the required correction from the syndrome~\cite{vittal2023astrea, wu2023fusionblossomfastmwpm}. The fraction of incorrect predictions is summarized by the probability of logical failure or LER. If the physical error rate is \textit{below threshold}, the LER decreases exponentially as the logical qubit is made larger, i.e., with increasing code distance,~$d$.

Although QEC handles physical errors using discrete detection events, the physical error rate depends on analog control parameters, such as pulse amplitude and drive frequencies, in addition to intrinsic noise such as decoherence. The optimal parameters for control naturally drift over time, raising physical error rates and, in turn, the LER. 
Drift can be addressed in two complementary ways: updating the decoder to reflect the change in the error model~\cite{wang2023dgr, sivak2024optimization}, or adjusting the controls to directly identify their new optimum. In this paper, we focus on runtime calibration of control parameters, using gate error amplification to efficiently minimize physical error rates by tracking the signal from firing detectors~\cite{gong2026provably, sivak2026reinforcement}. 

\begin{figure}[t]
    \centering
    \includegraphics[width=0.8\columnwidth]{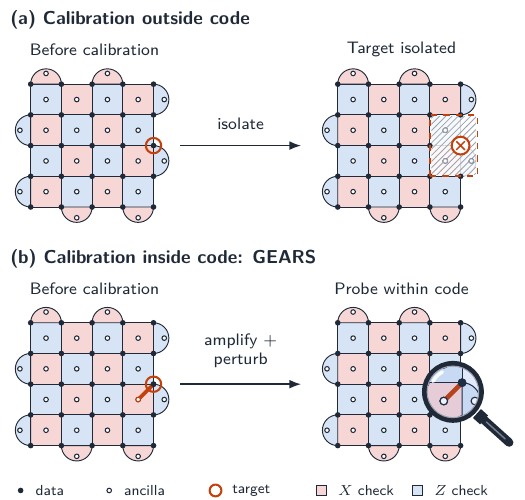}
    \caption{(a) Calibration outside the code requires isolating target qubits, which temporarily reduces logical protection. (b) GEARS probes selected gates in place by repeating them with small control perturbations. The controller uses the changes in local detection rates to update each gate's controls.}
    \label{fig:isolate-comparison}
\end{figure}

Traditionally, runtime calibration of a control parameter requires taking the affected set of qubits offline, as calibration disrupts quantum states. One such approach is to transfer the encoded logical information to a separate set of calibrated physical qubits; this pays qubit and execution-time overhead~\cite{fang2025caliqec}. Another approach is code deformation, which isolates offline qubits by modifying the QEC code (Fig.~\ref{fig:isolate-comparison}a); however, this temporarily weakens logical protection~\cite{fang2025caliqec}. Recent alternatives use runtime-generated metadata to recalibrate without taking qubits offline (Fig.~\ref{fig:isolate-comparison}b). Work by Sivak et al.~\cite{sivak2026reinforcement} proposed a way to use detector firing rates to estimate gradients that redirect the control parameters back to their optima. The parameters are perturbed about the current operating point, modifying the physical errors and detector firing frequency. An optimizer, in this case a reinforcement learning (RL) agent, estimates a policy gradient from changes in detector firing rates and uses it to update the controls. Thereafter, Gong et al.~\cite{gong2026provably} proved that the detector firing rate is a valid objective for runtime calibration of a large class of drift errors. Their optimizer, a locality-aware SPSA (LSPSA) algorithm, leverages the sparse relationship between control parameters and device errors to obtain gradients that, in their simulations, supported calibration performance comparable to or better than that of their RL baseline~\cite{sivak2026reinforcement}. \emph{In any case, successful runtime calibration depends on high-quality gradients.}

Gradient quality can deteriorate from device or statistical noise. These effects can be mitigated using: \textcircled{A} \textbf{Increased Number of Shots: } More shots yield a more accurate estimate of the expected detector firing rate at a specific control configuration. This yields higher-quality gradients, albeit at a higher shot cost. Sampling more shots costs more time and delays calibration feedback. \textcircled{B} \textbf{Larger Control Perturbations: } A larger perturbation leads to a larger deviation in detector firing rates and more pronounced gradients, but this significantly increases the physical error rate of all affected operations, compromising QEC performance. Ref.~\cite{sivak2026reinforcement} calls this added error \emph{exploration noise}. Perturbations that are too large can even move the system out of the locally convex region, invalidating some convergence assumptions~\cite{gong2026provably}. 

In this context, we propose \emph{GEARS: \underline{G}ate-\underline{E}rror \underline{A}mplification for \underline{R}ecalibration with \underline{S}yndromes}. GEARS keeps every qubit in the code during calibration: it amplifies and perturbs selected gates in place and uses the resulting detection rates to guide the update. GEARS enhances gradients by performing an $\ell$-fold amplification of the errors of selected physical gates in the syndrome-extraction circuit~(Fig.~\ref{fig:amplification-mechanism}a), reducing the need for more shots or larger perturbations. 
We use the fact that quantum gates under consideration are self-inverse operations and always choose $\ell$ to be an odd integer; this ensures logical correctness: for example, for the two-qubit CZ gate, $\text{CZ}^\ell = \text{CZ}\cdot(\text{CZ} \cdot \text{CZ})^{\frac{\ell-1}{2}} = \text{CZ}$ because $\text{CZ}^2 = I$. 

Gate repetition amplifies control errors, making local detection rates more sensitive to changes in the control parameters. Amplification of a single gate increases the average detection rate by an amount proportional to approximately $\ell^2$ (Fig.~\ref{fig:amplification-mechanism}b), while the injected logical-error cost is still exponentially suppressed with increasing code distance (Fig.~\ref{fig:amplification-mechanism}c, details in Sec.~\ref{amplfied-gradient}). 
GEARS uses \emph{dilution} to control the spacing between simultaneously probed gates, balancing the time cost of serialization against the logical cost of probing gates in parallel. We use surface code simulations to quantify the logical cost of gradient acquisition and evaluate the performance of GEARS when correcting static miscalibration over a fixed execution window and for tracking time-dependent drift.

Our main contributions are as follows:
\begin{enumerate}
    \item We introduce GEARS, a protocol that amplifies gate errors on selected gates during syndrome extraction to estimate control gradients. At $\ell=3$, GEARS requires 8$\times$ fewer measurements in order to achieve the same gradient-estimation accuracy as the unamplified case (Sec.~\ref{subsec:eval-acquisition}).
    \item In surface-code simulations with closed-loop calibration, we show GEARS at $\ell=3$ reduces excess logical failures by 31\%--40\% relative to calibration without amplification across the evaluated execution windows (Sec.~\ref{subsec:eval-fixed-window}). 
    \item Compared to the baseline, GEARS can track significantly faster drift. Under time-dependent drift, GEARS at $\ell=3$ reduces expected logical failures up to a critical drift frequency approximately 1.9$\times$ higher than that of calibration without amplification (Sec.~\ref{subsec:eval-drift}).
\end{enumerate}

\section{Background}\label{sec:background}

\subsection{Surface Codes}

The surface code is a widely used QEC approach because it only requires local interactions and supports a high error threshold~\cite{Fowler_2012}. Protecting quantum information requires correcting both bit flips ($X$) and phase flips ($Z$). The surface code detects these errors through repeated measurements of local $X$- and $Z$-type parity checks, or stabilizers. Ancilla qubits interact with nearby data qubits to measure these checks without directly measuring the encoded logical information~\cite{Fowler_2012}. A detector is defined by a set of measurements whose combined parity is deterministic in the absence of errors~\cite{sivak2026reinforcement, gidney2021stim}. A deviation from this expected parity constitutes a detection event or detector firing. During repeated QEC rounds, detectors check the parity of consecutive measurements of the same stabilizer to identify changes in its outcome. The decoder uses the resulting pattern across rounds to identify and correct physical errors.

The code distance $d$ quantifies the error correction capabilities of the surface code instance. Increasing $d$ suppresses logical failures exponentially---provided the physical error rate is below the error correction threshold~\cite{Fowler_2012}. Maintaining this protection therefore requires sufficiently low physical error rates, motivating continued calibration during computation.

\subsection{In Situ Calibration}

Quantum gates rely on analog controls whose calibrated settings can become inaccurate as device parameters and environmental conditions drift. Such drift is ubiquitous across quantum hardware platforms, including superconducting circuits~\cite{sivak2026reinforcement, superconductingdrift}, trapped ions~\cite{proctor_detecting_2020}, neutral atoms~\cite{evered_high-fidelity_2023}, semiconductor spins~\cite{capannelli_tracking_2025}, solid-state defects~\cite{solidstatedrift}, etc. Robust control can mitigate the effects of drift by reducing sensitivity to parameter changes~\cite{poggi2024universally, PhysRevApplied.17.014036, kamen2026comparing}, but recalibration remains necessary over longer intervals.

Conventional calibration uses dedicated state-preparation and measurement procedures that disturb quantum information. Applying these procedures during computation requires transferring the encoded state elsewhere or deforming the code to isolate the qubits being calibrated [Fig.~\ref{fig:isolate-comparison}(a)]~\cite{fang2025caliqec}. State transfer naturally incurs both qubit and execution-time overhead. Alternatively, code deformation provides this isolation by changing parity checks around unavailable qubits, allowing the remaining qubits to protect the encoded state~\cite{vuillot2019code, lin2024codesign}. Isolating qubits for calibration can reduce the code distance, which can be restored by expanding the code with additional physical qubits~\cite{fang2025caliqec}. Both approaches require spare physical qubits during calibration, either to hold the transferred logical state or to expand the patch to preserve the code distance.

Syndrome-based calibration avoids this qubit overhead by reusing the stabilizer measurements already performed during QEC to update controls, leaving the code intact [Fig.~\ref{fig:isolate-comparison}(b)]~\cite{kelly2016scalable}. Small control perturbations change detector statistics, providing feedback for adjustments that reduce the detector firing rate~\cite{sivak2026reinforcement}. Exploiting the local dependence of detector responses on control parameters allows for calibration at scale, with convergence guarantees that remain independent of code distance~\cite{sivak2026reinforcement, gong2026provably}.

The detection rate $C$, averaged across all detectors in the code, provides a surrogate objective for minimizing the logical-failure probability, $p_L$. Below threshold, logical failures become exponentially rare with increasing code distance, increasing the number of samples needed to estimate $p_L$ accurately; critically, $C$ does not shrink with code distance [Fig.~\ref{fig:amplification-mechanism}(b)-(c)]. 
For surface codes well below threshold, the gradients satisfy the approximate relation
\begin{align}
    \nabla_{\boldsymbol{\theta}}\log p_L
    \approx
    \frac{d+1}{2}\,
    \nabla_{\boldsymbol{\theta}}\log C,
\end{align}
where $\boldsymbol{\theta}$ denotes the control parameters and $\nabla_{\boldsymbol{\theta}}$ denotes the derivative with respect to $\boldsymbol{\theta}$. Approximate agreement has been observed experimentally for small control perturbations, supporting the use of detector statistics to guide calibration~\cite{sivak2026reinforcement}.

\subsection{Gate-Error Amplification}
\label{subsec:background-amplification}

Gate repetition can make small control errors easier to detect by allowing their effects to accumulate coherently before measurement. This idea also appears in quantum metrology, where repeated application of a parameter-dependent operation increases sensitivity to that parameter~\cite{giovannetti2006quantum}. Offline gate-calibration protocols leverage repetition to resolve small coherent errors and isolate specific error components~\cite{tornow2022minimum,gross2024characterizing}. Deliberate noise amplification also appears in zero-noise extrapolation (ZNE), an error mitigation technique that runs a quantum algorithm at several noise strengths to back out a zero-noise estimate of the algorithm outcome~\cite{temme2017error, zne1}. In ZNE, noise is increased by inserting gates while preserving the ideal computation~\cite{Giurgica_Tiron_2020}. Similarly, GEARS applies gate-error amplification within syndrome-extraction circuits to strengthen the local detector response to control perturbations while preserving the circuit's function. GEARS connects the sensitivity enhancement of repeated coherent operations with the syndrome-based feedback used for in situ calibration.

\section{GEARS: Design and Methodology}\label{sec:methodology}

\begin{figure*}[t]
    \centering
    \includegraphics[width=0.777\textwidth]{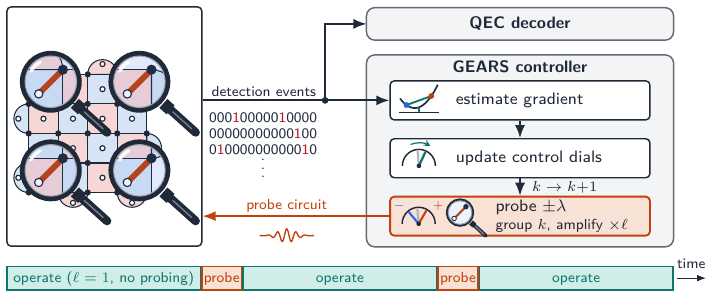}
    \caption{GEARS calibration workflow. Syndrome detection events are used for decoding and calibration. During a probing interval, GEARS selects spatially separated CZ gates, amplifies their control errors through gate repetition, and measures the response to positive and negative control perturbations. The controller uses these measurements to update gate controls before returning to unmodified error-correction operation.}
    \label{fig:gears-procedure}
\end{figure*}

GEARS is an in situ calibration mechanism that uses measurements already produced by QEC to adjust drifting gate controls. Its key idea is to temporarily amplify the effect of small control errors---making them easier to measure---without suspending computation. We first describe the calibration loop and explain how amplification improves gradient estimates. We then present the schedule for probing multiple gates in parallel, followed by the metrics and simulation model used to quantify calibration overhead.

\subsection{The GEARS Calibration Loop}\label{subsec:gears-calibration}

We target the CZ gates used in surface-code syndrome extraction. These gates contribute substantially to the error budget of superconducting QEC systems; even after initial calibration, their controls require adjustment as hardware conditions drift~\cite{google2025quantum}. Rather than removing qubits from the code to recalibrate them, GEARS performs short probing intervals within ongoing syndrome-extraction circuits (Fig.~\ref{fig:gears-procedure}). During each interval, GEARS perturbs the control parameters of each selected gate in the positive and negative directions. The total perturbation in either direction has magnitude $\lambda$, which we call the \emph{effective probe strength}. Outside these intervals, all gates execute with their current controls without amplification.

\textbf{Amplify a selected gate.} While probing, GEARS replaces a selected CZ gate with $\ell$ consecutive CZ gates, which we call an \emph{amplified block}. Since two ideal CZ gates cancel, an odd number of repetitions implements the same ideal operation as a single CZ; hence, $\ell$ is restricted to odd values, with $\ell=1$ the unamplified baseline. The repeated gates preserve the intended stabilizer circuit while allowing systematic control errors to accumulate in the block.

\textbf{Measure the control response.} Each CZ gate has several tunable control parameters. For a selected group of gates, GEARS probes one parameter at a time, but probes that parameter on all gates in the group concurrently. The controller runs two batches of QEC memory shots, each of size $B$. For each gate, it applies a small perturbation with a randomly chosen sign in the first batch and the opposite sign in the second. Randomizing signs across gates cancels cross-terms in expectation and helps isolate the response of each gate's control parameter~\cite{spall1992multivariate,duchi2015optimal,gong2026provably}.

The controller estimates the direction in which each parameter should change from the difference in local detector firing rates between the two batches. After probing all parameters, it updates the selected gates using the Adam optimizer~\cite{kingma2014adam} and proceeds to the next group, or ends the probing interval according to the calibration schedule. Gates outside the selected group continue to execute without amplification or control perturbations.

\subsection{Gradient Estimation with Amplification}\label{amplfied-gradient}

GEARS uses \emph{detection events}---changes in stabilizer measurement outcomes that indicate possible errors---as feedback for calibration~\cite{kelly2016scalable,sivak2026reinforcement,gong2026provably}. A gate's control gradient describes how the average detection rate changes when a control parameter is adjusted. A positive-versus-negative probe estimates this gradient without requiring direct access to the underlying physical error parameters.

\textbf{Use local detector feedback.} A CZ gate affects only a small neighborhood of detectors in a surface-code syndrome extraction circuit. For each gate $j$, GEARS tracks only the detectors that respond to faults at that gate, denoted $\mathcal D_j$, instead of collecting a chip-wide score for every control update. We define the local score, $c_j$, as the sum of detection-event rates in this neighborhood, divided by the total number of detectors in the code. Detectors outside $\mathcal D_j$ are insensitive to gate $j$'s controls, so they do not contribute to its gradient. This locality is the basis of scalable syndrome-based calibration for all quantum low-density parity check (qLDPC) codes, so named because their checks and qubit connectivity have bounded degree (the surface code is a simple case of qLDPC code)~\cite{gong2026provably}.

\textbf{Normalize the amplified response.} Each of the $\ell$ CZ copies receives a perturbation of magnitude $\lambda/\ell$, keeping the effective probe strength fixed at $\lambda$. Near the calibrated operating point, detector firing rates have an approximately quadratic dependence on control offsets, as observed experimentally~\cite{sivak2026reinforcement} and proven for circuits with Pauli-twirled noise~\cite{gong2026provably}. For the CZ control-error model considered here~\cite{gross2024characterizing}, the error unitary commutes with the ideal gate. For a given control offset, repeating the gate $\ell$ times therefore increases the accumulated offset by a factor of $\ell$. The slope of this quadratic response is proportional to the accumulated offset, so the slope at the amplified operating point scales approximately linearly with $\ell$. At the same effective probe strength, the difference between the positive and negative probe responses is therefore approximately $\ell$ times larger with amplification than without amplification (see Appendix~\ref{app:amplification-normalization} for the mathematical details).

To measure this difference, GEARS randomly increases or decreases each gate's selected control parameter in the first batch. The second batch reverses these perturbations. For gate $j$, we use $c_+$ for the measured local score over $\mathcal D_j$ when the parameter is increased and $c_-$ when it is decreased (where we suppress the subscript $j$ for simplicity). The gradient estimate is
\begin{equation}
    \hat g = \frac{c_+-c_-}{2\lambda\ell}.
    \label{eq:normalized-gradient}
\end{equation}
The numerator measures how much the local detector firing rate changes when the control is perturbed. The denominator accounts for the positive-to-negative probe separation ($2\lambda$) and the extra sensitivity produced by gate repetition ($\ell$). For $\ell=1$, we recreate the Locality-aware Simultaneous Perturbation Stochastic Approximation (LSPSA) estimator used in Ref.~\cite{gong2026provably}. GEARS improves the measurement efficiency of this estimator by using gate-error amplification to increase the detector response to small control offsets.

This larger difference between the probe responses enables us to estimate the same gradient with significantly lower variance after normalization. Each local detection rate is estimated from a batch of $B$ memory shots per probe direction. In this case, the gradient variance scales as
\begin{equation}
    \operatorname{Var}(\hat g) \propto \frac{1}{B\lambda^2\ell^2}.
    \label{eq:gradient-variance-scaling}
\end{equation}

This captures the main measurement benefit: at fixed effective probe strength, $\lambda$, increasing the repetition count by a factor of $\ell = 3$ can reduce the shots needed for the same gradient variance by $\sim 9 \times$. Appendix~\ref{app:sampling-variance} gives the full derivation.

Increasing the probe strength, $\lambda$, also reduces sampling variance; however, this introduces larger control perturbations that can increase the logical-error probability per shot because we deviate significantly from the original operating point. It can also increase the bias of the gradient estimate [Fig.~\ref{fig:gradient-bias}]. GEARS circumvents these issues by increasing measurement sensitivity at a fixed effective value of $\lambda$. 

Amplification incurs alternate costs. Repetition magnifies existing control errors and adds stochastic gate errors. Qubits outside the amplified gates also accumulate idling errors during the extra repetitions, which might increase the logical error rate. Hence, the central design question is: \emph{Do the measurement savings outweigh the higher logical-error cost per amplified shot?} We evaluate this tradeoff in Sec.~\ref{subsec:eval-acquisition}.

\subsection{Spatially Parallel Probing}\label{subsec:spatial-probing}

Probing gates one at a time would make calibration latency grow with the number of CZ gates. GEARS instead exploits the local response of detectors to probe many gates in the same measurement batch. Parallel probes must satisfy two constraints: their detector responses should not overlap, so that the gradient estimates remain distinguishable~\cite{kelly2016scalable,sivak2026reinforcement,gong2026provably}, and their induced errors should be sufficiently separated to avoid concentrating faults in one region of the code.

GEARS constructs probing groups offline from the syndrome-extraction circuit. First, it separates CZ gates by their temporal layer; gates in different layers are never probed in the same batch. For each gate within a layer, it then propagates representative Pauli faults placed immediately before and after the gate through an otherwise noiseless circuit. This identifies two \emph{footprints}: the detectors that can respond to the fault and the data qubits through which the fault can propagate.

Two gates may share a probing group only if their detector footprints do not overlap and their data-qubit footprints are separated by more than a configurable \emph{dilution radius}, $R$. The dilution radius is measured on the two-dimensional code layout using Chebyshev distance (the larger of the horizontal and vertical coordinate separations, minimized over qubit pairs in the two footprints). The radius lets the controller trade probing parallelism against the spatial concentration of probing-induced errors.

We implement grouping as a graph-coloring problem. Within each temporal layer, a vertex represents a CZ gate, and an edge connects two gates if their detector footprints overlap or their data-qubit footprints violate the separation constraint. We color this conflict graph using DSATUR~\cite{brelaz1979new, gong2026provably}; gates of the same color form a probing group. For a fixed dilution radius, the number of groups is governed by local conflicts rather than the total number of gates or the code distance. Increasing $R$ may require more groups, increasing measurement overhead while providing greater spatial separation. Appendix~\ref{app:spatial-probing} describes the detector footprint and dilution radius in greater detail.

\subsection{Measurement Requirement and Logical Cost}\label{sec:cost}

We evaluate gradient acquisition using two metrics: the number of measurements needed for a sufficiently accurate control update, and the expected number of logical failures while estimating the gradient. Ideally, the number of expected failures is much less than one.

\textbf{Gradient accuracy.} We quantify the accuracy of the estimated control gradient using the signal-to-noise ratio (SNR),
\begin{equation}
    \mathrm{SNR} = \frac{g_{\mathrm{rms}}}{e_{\mathrm{rms}}}.
\end{equation}
The numerator $g_{\mathrm{rms}}$ is the root-mean-square magnitude of reference gradients across all gates and control parameters. To obtain a reference gradient free of sampling noise, we vary one control parameter at a time by a very small amount ($\pm\text{5e-5}$ radians) and take a central difference using the local contributions $c_\pm$ computed directly from the unamplified circuit model's exact detector firing probabilities. The denominator $e_{\mathrm{rms}}$ is the root-mean-square error of the gradient estimate relative to the reference gradient. It includes both sampling noise and systematic bias (see Appendix~\ref{app:gradient-accuracy}).

If each gate has $m$ control parameters and the schedule has $W$ probing groups, one complete gradient measurement uses $N=2BmW$ memory shots, where the factor of two comes from the two polarities per parameter per group. Importantly, the number of gates \emph{within} a probing group does not multiply this count, because those gates are measured in parallel. For each amplification depth $\ell$, we find the smallest batch size $B$ that meets the target SNR, and call the resulting shot count $N_\ell$.

\textbf{Logical cost.} Let $\overline p_L^{(\ell)}$ be the per-shot LER averaged over probing groups, control parameters, and probe polarities. We define the acquisition cost as $K_\ell=N_\ell\,\overline p_L^{(\ell)}$: the expected number of logical failures incurred while acquiring a gradient at the target accuracy. This includes failures caused by background noise as well as probing. To evaluate this cost, we decode syndromes using minimum-weight perfect matching with PyMatching~\cite{Higgott2025sparseblossom}. In the closed loop experiments, decoder weights are fixed using the noise model of the calibrated circuit without probing. Updating the decoder weights to account for probing could further reduce the logical cost of gradient acquisition, but we leave this to future work and focus on isolating the effect of amplification. For calibration to coexist with a fault-tolerant computation, the desired regime is $K_\ell\ll1$. Section~\ref{subsec:eval-acquisition} evaluates how amplification changes the shot count and logical cost, and how these effects scale with code distance.

\subsection{CZ Error Model and Circuit Simulation}\label{subsec:cz-error-model}

GEARS is designed to exploit coherent error amplification across repeated CZ gates; however, simulating the entire surface-code circuit with general coherent noise is expensive, while efficiently modeling the error contribution of each gate as an independent stochastic process disregards the essential coherence. Therefore, we first compose the imperfect gates within an amplified block to accumulate errors, and afterward we convert the combined error into a stochastic error channel to preserve simulation efficiency.

We model CZ control errors using five independently adjustable coordinates, following the parametrization used to characterize CZ gates on Google's Sycamore processor~\cite{gross2024characterizing}. They describe controlled-phase error ($\phi$), two swap-error components ($x$ and $y$), and relative and common single-qubit phase errors ($\zeta$ and $\gamma$). We treat these coordinates as proxies for tunable hardware controls. At their calibrated values, coherent error vanishes in the model, although stochastic background errors remain.

For each probing configuration, we multiply the $\ell$ imperfect CZ operations, factor out the ideal CZ to obtain the residual error unitary, and apply the Pauli-twirling approximation to this residual error~\cite{emerson2007symmetrized,geller2013efficient}. Applying twirling \emph{after} gate composition preserves the effect of coherent accumulation in the resulting Pauli error probabilities. We can then evaluate large syndrome-extraction circuits efficiently in Stim~\cite{gidney2021stim}. The full unitary model and channel construction appear in Appendix~\ref{app:coherent-error-model}. In a hardware implementation, randomized compiling with sufficiently accurate single-qubit gates can tailor coherent errors into stochastic Pauli noise~\cite{wallman2016noise}. Measurements of a CZ gate on a superconducting processor showed good agreement with this noise model under randomized compiling~\cite{hashim2023benchmarking}. Moreover, QEC naturally decoheres coherent errors, effectively approximating a randomized compiler~\cite{beale2018quantum}.

We also include stochastic background noise using the SI1000 model~\cite{gidney2021fault}. Every CZ copy, including copies added by amplification, experiences two-qubit depolarizing noise with probability $p=10^{-3}$. Single-qubit gates and idle time steps have depolarizing probability $p/10$; reset and measurement errors have probabilities $2p$ and $5p$, respectively; and qubits idling during reset or measurement receive an additional depolarizing channel with probability $2p$. We assume a stabilizer round duration of $1.1\,\mu\mathrm{s}$ without amplification and a CZ gate duration of $42\,\mathrm{ns}$, following Ref.~\cite{google2025quantum}. Each additional amplification layer extends the round by $42\,\mathrm{ns}$. Qubits idling during these temporal layers receive depolarizing noise with probability $p/10$ per layer, following the SI1000 model~\cite{gidney2021fault}. Together, these models account for both the stronger calibration signal and the noise overhead of amplification.
\section{Evaluation}\label{sec:evaluation}

\subsection{Gradient estimation SNR}
\label{subsec:eval-accuracy}

\begin{figure}[t]
    \centering
    \includegraphics[width=0.8\columnwidth]{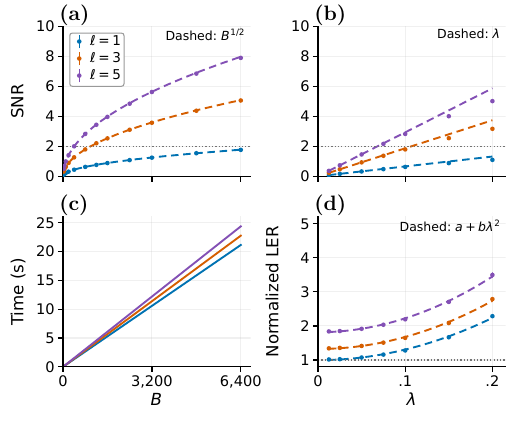}
    \caption{Gradient estimation accuracy and probing cost. The upper row shows gradient SNR as a function of (a) batch size $B$ at fixed probe strength $\lambda=0.1$ and (b) probe strength $\lambda$ at $B=800$ shots per polarity. Dashed curves show the predicted $\sqrt{B}$ and $\lambda$ scaling, respectively, normalized to the first point in each series. Error bars indicate 95\% confidence intervals. Panel (c) shows the wall clock time to estimate a gradient as a function of batch size $B$. Panel (d) shows the mean per-shot LER versus probe strength, normalized to the LER without probing at the same control offset. Dashed curves in (d) show quadratic fits.
}
    \label{fig:snr-scaling}
\end{figure}

We first evaluate whether GEARS can use fewer measurements or weaker probes and match the gradient accuracy of calibration without amplification. When sampling noise is the dominant source of error, $e_{\mathrm{rms}}^2$ is approximately equal to the average sampling variance, so Eq.~\eqref{eq:gradient-variance-scaling} gives $e_{\mathrm{rms}}\propto 1/(\lambda\ell\sqrt{B})$. Since the reference gradient magnitude $g_{\mathrm{rms}}$ is fixed, substituting this scaling into $\mathrm{SNR}=g_{\mathrm{rms}}/e_{\mathrm{rms}}$ from Eq.~(4) gives
\begin{align}
    \operatorname{SNR}\propto \lambda\ell\sqrt{B}.
     \label{eq:snr-scaling}
\end{align}

To validate this dependence, we use a distance-5 surface code with 15 stabilizer rounds per memory shot and a dilution radius $R=0$. For each CZ gate, we independently draw the five residual control angles from a standard normal distribution, then rescale the resulting vector to a Euclidean norm $s$. We set $s=0.02$ radians to study gradient estimation for a small control offset, which is expected when tracking drift at runtime after offline calibration has brought the controls close to the optimum~\cite{gong2026provably}.

Our results confirm that for small control offsets, amplification improves SNR at fixed batch size and probe strength, allowing the same accuracy with fewer shots or weaker probes. Figure~\ref{fig:snr-scaling}(a) shows the predicted $\sqrt{B}$ scaling at $\lambda=0.1$. Figure~\ref{fig:snr-scaling}(b) shows that SNR grows approximately linearly with probe strength through $\lambda=0.2$. At higher probe strengths, SNR falls below the linear prediction in Fig.~\ref{fig:snr-scaling}(b), reflecting changes in sampling variance at the two probe polarities (Appendix~\ref{app:sampling-variance}) and increasing estimator bias (Appendix~\ref{app:gradient-accuracy}). A stronger probe introduces more coherent error and can push the controls beyond the range where the local detection rate is approximately quadratic in the control offset. The mean gradient estimate can then differ systematically from the reference. This difference, measured by the relative RMS bias, increases with probe strength at all three amplification depths (see Appendix Fig.~\ref{fig:gradient-bias}). Figure~\ref{fig:snr-scaling}(c) shows the wall clock time to estimate a gradient. At fixed batch size, amplification adds only a small time overhead because the additional CZ repetitions are short compared with a full syndrome-extraction round ($42$ns vs 1.1$\mu$s). Figure~\ref{fig:snr-scaling}(d) shows that increasing probe strength raises the logical failure probability approximately quadratically over the range shown. If the batch size is held fixed, the SNR scaling is such that an unamplified probe must be $\ell$ times stronger to match the SNR at depth $\ell$, increasing the logical cost per shot. Alternatively, at fixed probe strength, the unamplified baseline would need to increase the batch size $B$ by a factor of $\ell^2$ to achieve the same SNR, increasing execution time. Rather than increasing the batch size or effective probe strength, GEARS magnifies the residual control offset---which is likely to be small during continuous calibration~\cite{gong2026provably}. Although we expect this amplified offset to be small, there are additional costs from stochastic CZ errors and idling errors introduced by gate repetition. We account for these tradeoffs in the numerical experiments that follow.

\subsection{Spatial dilution and logical cost}
\label{subsec:eval-dilution}

\begin{figure}[t!]
    \centering
    \includegraphics[width=0.80\columnwidth]{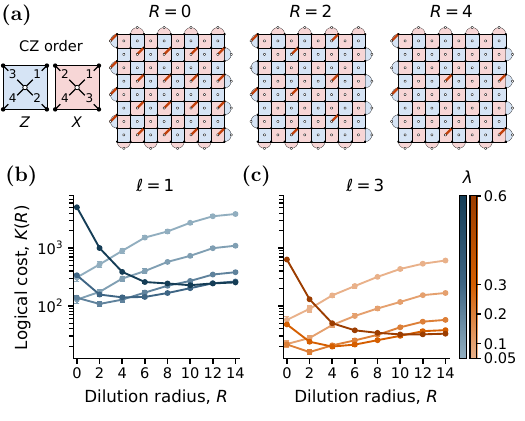}
    \caption{Spatial dilution and the logical cost of gradient estimation. (a) Numbers give the CZ gate order for each check. For each dilution radius $R$, the red bonds mark the CZ gates probed together in the first group of layer~1. (b) Expected logical failures $K$ during gradient acquisition versus $R$ without amplification ($\ell=1$). (c) The same comparison with amplification ($\ell=3$). Darker shades indicate larger probe strengths $\lambda$. All curves use the same distance-9 circuit and control offset ($s=0.02$) at a target gradient SNR of 2. Error bars indicate 95\% confidence intervals.}
    \label{fig:dilution}
\end{figure}

We examine how spatial dilution affects the logical cost of gradient acquisition, where we set $\operatorname{SNR}_{\mathrm{target}}=2$. This requires the root-mean-square estimation error to be no greater than half the gradient's norm, i.e., $e_{\mathrm{rms}}\le g_{\mathrm{rms}}/2$. Fig.~\ref{fig:dilution} shows the simulation results of a distance-9 surface code with 27 rounds of syndrome extraction and a small residual control offset of $s=0.02$. We vary the dilution radius $R$ at probe strengths $\lambda\in\{0.05,0.1,0.2,0.3,0.6\}$, comparing calibration without amplification ($\ell=1$) with GEARS at $\ell=3$.

Increasing $R$ separates simultaneously probed gates into more groups, as illustrated in Fig.~\ref{fig:dilution}(a). This reduces the concentration of errors introduced by probing but requires additional measurement batches. Fig.~\ref{fig:dilution}(b) and Fig.~\ref{fig:dilution}(c) show the resulting logical acquisition cost, including measurements across all probing groups.

The tradeoff depends on probe strength. At $\lambda=0.05$, the additional measurement requirement outweighs the reduction in logical disturbance, and acquisition cost increases with $R$. At $\lambda=0.3$, increasing the radius from $R=0$ to $R=4$ reduces the cost by approximately 60\% for both amplification depths, indicating that stronger probing benefits from greater serialization. At $\lambda=0.6$, the lowest logical costs occur near $R=10$, at approximately twenty times lower than at $R=0$.  

Amplification lowers the estimated logical acquisition cost at every sampled combination of probe strength and dilution radius. At distance nine, the three intermediate probe strengths illustrate how the lowest-cost radius changes with probe strength, from $R=0$ at $\lambda=0.1$ to $R=2$ at $\lambda=0.2$ and $R=4$ at $\lambda=0.3$. These settings yield comparable costs for each amplification depth. Acquiring one gradient incurs about 108--140 expected logical failures without amplification and 16--22 with depth-three GEARS.

Even with amplification, a single gradient acquisition at distance nine still incurs tens of expected logical failures. We therefore examine how acquisition cost scales with code distance, with the eventual goal of bringing it well below one expected logical failure per gradient acquisition.

\vspace{-4pt}

\subsection{Logical cost of gradient acquisition}
\label{subsec:eval-acquisition}

\begin{figure}[t]
    \centering
    \includegraphics[width=0.8\columnwidth]{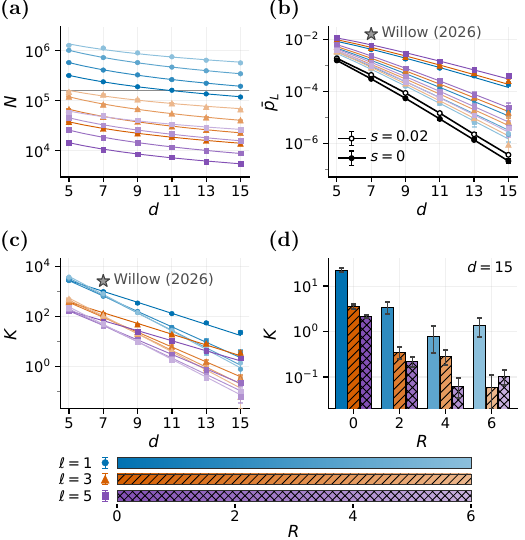}
    \caption{Amplification lowers the logical cost of gradient acquisition. At a target gradient SNR of 2, panels (a--c) show the required shot count $N$, mean per-shot LER $\bar{p}_L$, and expected logical failures $K=N\bar{p}_L$ versus code distance $d$. Amplification raises $\bar{p}_L$ but reduces $N$ by more, so $K$ is lower than without amplification at every simulated distance. Curves show fits in (a) and (b) and their product in (c). Panel (d) shows $K$ versus dilution radius $R$ at $d=15$. The gray line in (a) marks the typical budget of 160{,}000 shots per training epoch reported by Sivak et al.~\cite{sivak2026reinforcement}. The star in (b) shows the projected 21-round failure probability at their calibrated rate. The star in (c) shows the expected logical failures over that shot budget at the same rate, excluding exploration noise. Error bars indicate 95\% confidence intervals.}
    \label{fig:matched-gradient}
\end{figure}

For the distance sweep in Fig.~\ref{fig:matched-gradient}, we fix the probe strength at $\lambda=0.3$ and use the same targeted gradient SNR of 2 and residual control offset of $s=0.02$. We simulate distances $d=5$--$15$ with $3d$ stabilizer rounds per memory shot, comparing amplification depths $\ell\in\{1,3,5\}$ and dilution radii $R\in\{0,2,4,6\}$. Points show estimates at each distance. Colored curves fit $N\propto(3d)^{-\beta}$ and $\overline p_L\propto 3d\,e^{-\alpha d}$ in panels (a) and (b), respectively; their products give the curves in panel (c). The black reference lines linearly interpolate sampled values. For the star in Fig.~\ref{fig:matched-gradient}(b), we use the failure probability equation from Ref.~\cite{bausch2024alphaqubit} to convert the distance-7 per-cycle LER $\epsilon_L=7.72\times10^{-4}$ reported by Sivak et al.~\cite{sivak2026reinforcement} to an estimated per-shot LER $\bar{p}_L=[1-(1-2\epsilon_L)^{21}]/2\approx1.60\%$ for a 21-round memory shot, matching the shot length used in our distance-7 simulations.

Across the evaluated distances, amplification reduces the measurement requirement for achieving the target SNR [Fig.~\ref{fig:matched-gradient}(a)]. At the same distance and dilution radius, $\ell=3$ and $\ell=5$ require approximately eight and twenty times fewer shots, respectively, than calibration without amplification. These reductions are roughly consistent with the inverse-square dependence on amplification depth predicted by the SNR scaling. Larger dilution radii increase the total number of shots needed to measure the gradient because probing is serialized across more groups.

The black reference curves in Fig.~\ref{fig:matched-gradient}(b) show the per-shot LER without probing at the residual control offset ($s=0.02$) and at the calibrated point ($s=0$). At the same dilution radius and code distance, amplification generally increases the logical failure probability per shot [Fig.~\ref{fig:matched-gradient}(b)]. However, amplification reduces the required shot count enough to offset the higher per-shot LER, lowering the logical acquisition cost $K=N\overline p_L$ [Fig.~\ref{fig:matched-gradient}(c)]. 

Dilution trades a larger shot count for a lower $\bar{p}_L$, and this tradeoff improves with code distance. At $\ell=3$, increasing the radius from $R=0$ to $R=4$ raises the acquisition cost by 17\% at $d=5$ but lowers it by 59\% at $d=9$ and 92\% at $d=15$. Similarly, $R=6$ costs more than $R=0$ at $d=5$, where the added shots outweigh the lower $\bar{p}_L$, but costs less at larger distances.

Fig.~\ref{fig:matched-gradient}(d) summarizes the costs at $d=15$. At $R=4$, the estimated number of logical failures incurred during gradient acquisition decreases from $0.77$ without amplification to $0.28$ at $\ell=3$ and $0.063$ at $\ell=5$. The latter corresponds to an approximately twelvefold reduction in estimated logical acquisition cost relative to calibration without amplification. In contrast to the previous distance-9 result, the logical acquisition cost at distance-15 demonstrates gradient acquisition with less than one expected logical failure in the simulated regime---with lower acquisition costs using GEARS.

\subsection{Calibration over a fixed execution window}
\label{subsec:eval-fixed-window}

\begin{figure}[t]
    \centering
    \includegraphics[width=0.8\columnwidth]{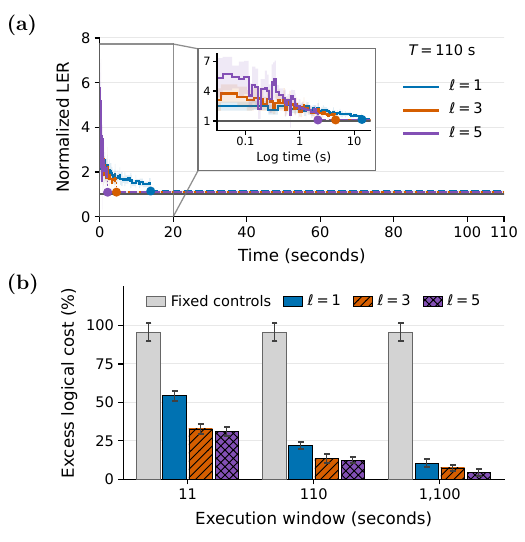}
    \caption{For a static control offset, GEARS has an upfront logical cost while probing, but faster calibration can reduce excess logical cost over the full execution window. (a) Expected logical failures per second during the 110-second window, normalized to the rate with calibrated controls and no probing. Circles mark the switch from probing (solid) to holding the controls fixed (dashed). (b) Excess logical cost as a percentage of the calibrated baseline (Appendix~\ref{app:batch-size-sensitivity}), with gray bars showing the cost of leaving the initial offset uncorrected. For each amplification depth and execution window, we choose the batch size and when to stop probing to minimize the mean logical cost across 16 runs. Shading and error bars show one standard deviation about the mean.}
    \label{fig:fixed-window}
\end{figure}

The preceding experiments measure the cost of acquiring an accurate gradient. We now evaluate whether the resulting control updates reduce logical failures over a complete execution window, accounting for both probing and subsequent operation. We use a distance-5 surface code because simulating full calibration trajectories is computationally expensive. Each memory shot comprises 15 stabilizer rounds. We use a weak probe ($\lambda=0.1$), which gave less than $2\%$ relative gradient bias at all three amplification depths at distance five. At distance five, increasing $R$ from 0 to 4 raised acquisition cost even for the stronger $\lambda=0.3$ probe with $\ell = 3$ GEARS (Fig.~\ref{fig:matched-gradient}). Since the weaker probe introduces less disturbance, we expect less benefit from dilution and use $R=0$ to avoid the extra measurement batches. With this schedule, we compare amplification depths $\ell\in\{1,3,5\}$ over execution windows of 11, 110, and 1100 seconds.

Each CZ gate starts with a control offset in a random direction. We set the offset magnitude to $s=0.0375$ radians to simulate a small, static miscalibration. We use the same 16 random seeds across all batch sizes and amplification depths, and we set the Adam learning rate to be $\eta=0.005$ at all three amplification depths.

For each depth and execution window, we vary how long calibration runs at each batch size. Once probing stops, the controls remain fixed for the rest of the window. We select the batch size and probing duration that minimize the expected number of logical failures over the full window, averaged over the same 16 random seeds. Appendix~\ref{app:batch-size-sensitivity} reports the batch size comparisons.
\begin{figure*}[!t]
    \centering
    \includegraphics[width=0.8\textwidth]{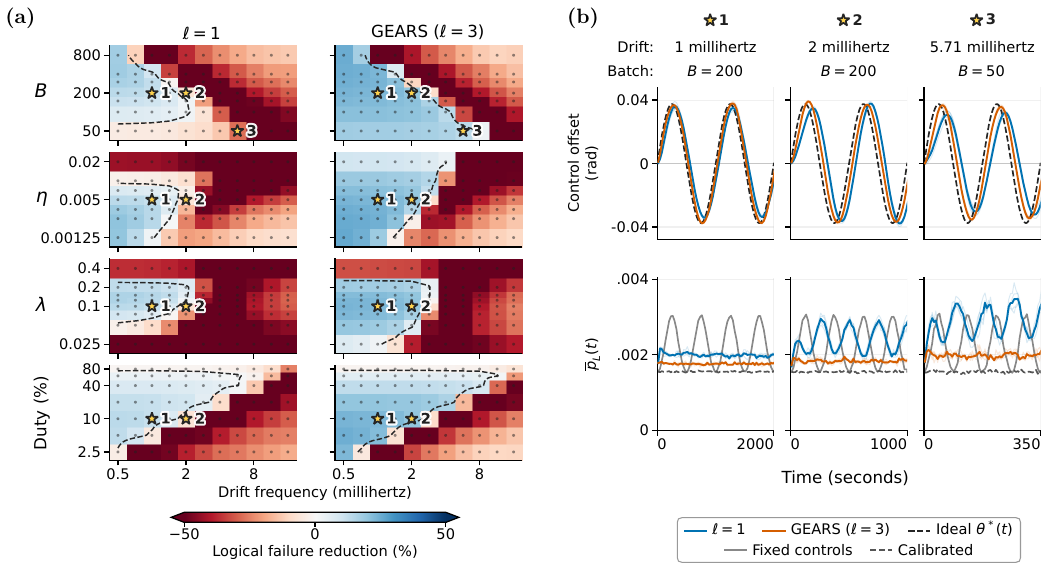}
    \caption{For sinusoidal drift, GEARS reduces logical failures over a larger range of frequencies. (a) Percentage reduction in logical failure probability per unit time relative to fixed controls, averaged over four seeds. The comparison includes both probing and ordinary operation. The four rows sweep batch size $B$, learning rate $\eta$, probe strength $\lambda$, and probing duty cycle, one at a time. Dashed contours mark zero improvement. (b) Representative trajectories for the three starred settings in (a). The upper plots show how closely the controls track the moving optimum along each gate's drift direction. The lower plots show how the probability of a logical failure changes over time.}
    \label{fig:drift-calibration}
\end{figure*}

Fig.~\ref{fig:fixed-window}(a) shows the optimal probing schedules for the 110-second window. Probing initially raises the logical failure rate, but the resulting control updates reduce errors during subsequent operation. The selected schedules balance this initial cost against the benefit of improved controls over the remaining window.

Fig.~\ref{fig:fixed-window}(b) reports the excess logical cost above perfectly calibrated operation, with error bars indicating one standard deviation over 16 random seeds. At 110 seconds, GEARS reduces excess logical cost by approximately 39\% at $\ell=3$ and 43\% at $\ell=5$, relative to calibration without amplification. At $\ell=5$, the reductions are approximately 43\% and 56\% for the 11- and 1100-second execution windows, respectively. These results show that savings in gradient acquisition translate to lower end-to-end logical cost after accounting for probing disturbance and residual errors.

\subsection{Tracking time-dependent drift}
\label{subsec:eval-drift}

In the fixed-window experiment, calibration corrects a static control offset. As in Ref.~\cite{sivak2026reinforcement}, we now let the optimal controls drift and estimate the highest drift frequency at which each calibration method still reduces expected logical failures per unit time. We evaluate drift tracking using a distance-5 surface code with 15 rounds per memory shot, dilution radius $R=0$, and probe strength $\lambda=0.1$. For each CZ gate $j$, the optimal controls drift according to
\begin{align}
    \boldsymbol{\theta}_j^*(t)
    =
    s_{\mathrm{drift}}\boldsymbol{u}_j\sin(2\pi f t),
\end{align}
where $\boldsymbol{u}_j$ is a fixed random unit vector in the five-dimensional control space and $f$ is the drift frequency. The directions are drawn independently for each gate, while all gates share the same frequency and phase. We set $s_{\mathrm{drift}}=0.0375$ radians, matching the initial offset magnitude in the static calibration experiment. Each experiment starts with zero control offset and runs for four drift periods. We compare GEARS at $\ell=3$ with calibration without amplification at $\ell=1$. We use $B=200$, $\eta=0.005$, $\lambda=0.1$, and a 10\% probing duty cycle as a common reference configuration. At this setting, both GEARS and unamplified calibration reduce logical failures relative to fixed controls at 1~mHz. To keep the number of simulated calibration trajectories manageable, we sweep each parameter separately at each drift frequency, holding the other three at their reference values. Each individual probe step alternates with ordinary operation; at 10\% duty, ordinary operation lasts nine times the amount of time it takes to get one gradient estimate. The Adam optimizer updates the controls after each probing group, and the controls remain fixed between probing sweeps while the optimum drifts. We repeat each setting with four random seeds, keeping the gate drift directions fixed across all runs.

At each drift frequency, we compare GEARS with unamplified calibration by varying batch size, learning rate, probe strength, and probing duty cycle one at a time [Fig.~\ref{fig:drift-calibration}(a)]. We average the LER over the full execution time, including both probing and ordinary operation. Colors show the percentage reduction in this rate relative to holding the initial controls fixed. Varying the probe strength tests whether stronger unamplified probes can recover the tracking benefit of GEARS. In the probe strength sweep, GEARS achieves a larger mean reduction in logical failures at every tested setting where both methods improve over fixed controls. 

The critical drift frequency marks where calibration no longer reduces expected logical failures relative to fixed control. For each parameter sweep in Fig.~\ref{fig:drift-calibration}(a), we represent this with the dashed contours to mark where calibration and fixed controls yield the same LER.  When considering this critical drift frequency,  we note how increasing the probing duty cycle allows for calibration at faster drift, but the added probing errors gradually decrease the improvement---at 80\% duty cycle, both methods increase expected logical failures across all tested frequencies. Across the parameter sweeps in Fig.~\ref{fig:drift-calibration}(a), we next select the set of parameters at each frequency where the average reduction in expected logical failures is the largest. We estimate the critical drift frequency by linearly interpolating these values to find where the mean reduction crosses zero. This estimated critical drift frequency is at 6.44~mHz without amplification and increases to 12.17~mHz with GEARS, an approximately 1.9$\times$ improvement.

Figure~\ref{fig:drift-calibration}(b) shows how the controls track the moving optimum at the three representative examples marked by stars in Fig.~\ref{fig:drift-calibration}(a). At 1 millihertz with the reference configuration, GEARS follows the optimum more closely than unamplified calibration. Over four drift periods, expected logical failures per unit time decrease by approximately 23\% with GEARS and 12\% with unamplified calibration. At 2 millihertz with the same settings, GEARS reduces failures by 20\%, while unamplified calibration increases them by 7\%. At 5.7 millihertz, we reduce the batch size to $B=50$ and keep the other settings fixed. GEARS reduces failures by 13\% at this frequency, while unamplified calibration increases them by 28\%. These examples show how GEARS can track the optimal controls more closely over a larger range of frequencies. 

\section{Discussion}\label{sec:discussion}

Our simulations use a superconducting-inspired circuit-noise model. Implementing GEARS on hardware would test the predicted gains in gradient estimation under experimental noise and latency constraints. Sequences inspired by Matrix-Element Amplification using Dynamical Decoupling (MEADD) could extend sensitivity to error components that do not accumulate constructively with just gate repetition. MEADD selectively amplifies non-commuting coherent errors by interleaving gates with single qubit dynamical decoupling pulses~\cite{gross2024characterizing}. Pulse stretching, already demonstrated for noise amplification in zero-noise extrapolation~\cite{kandala2019error}, could offer another way to integrate amplification with syndrome-based calibration.

Extending GEARS to neutral-atom and trapped-ion processors would allow evaluation under different noise mechanisms and timing constraints. In neutral-atom processors, measurements and their associated overhead can take much longer than gate operations~\cite{bluvstein2026fault}. This makes increasing the information obtained per measurement particularly valuable. Additional gate repetitions could strengthen calibration feedback at a small time cost, reducing the number of slow measurement cycles that are needed to acquire a gradient.

The right probing schedule depends on the specific hardware and the timescales of control drift. In our simulations of sinusoidal drift, a higher probing duty cycle allows both GEARS and unamplified calibration to track faster drift but gives a smaller reduction in average LER at lower frequencies. Calibration feedback acts like a filter on control drift: in Ref.~\cite{sivak2026reinforcement}, the ratio of LER power spectral densities with and without feedback gives an empirical filter function (see their Fig.~S5). A measured control-noise spectrum could therefore guide the duty cycle as well as the batch size, learning rate, and probe strength, where the optimal filter function depends on the control-noise spectrum. The drift simulations in Fig.~\ref{fig:drift-calibration} took an estimated $\sim$4000 CPU core-hours, and an exhaustive four-parameter sweep would take roughly 100 times that, so the reasonably tractable approach is to tune these parameters iteratively. Instead of a fixed duty cycle, detection rates could also be used to trigger probing adaptively (Appendix~\ref{app:rho}), globally or per group, though the time to resolve a change in detection rate might limit how fast the drift can be tracked. 

\section{Conclusion}\label{sec:conclusion}

GEARS uses gate-error amplification to improve the measurement efficiency of syndrome-based calibration. Amplification strengthens the detector response to control perturbations, reducing the measurements required to estimate gradients at a specified SNR. Circuit-level simulations show that this reduction can outweigh the additional logical failures introduced by probing. In closed-loop experiments, GEARS lowers the total logical cost for static offsets as well as faster time-dependent drift. These results demonstrate that deliberately amplifying physical control errors can reduce the overall logical cost of identifying and correcting them.

\section*{Acknowledgements}\label{sec:ack}

\noindent\textbf{AI disclosure.} Claude Fable 5.0--5.2 and GPT-6 Astra assisted with code development and manuscript preparation. The authors take full responsibility for the contents of this manuscript.

This work is funded in part by the STAQ project under award NSF Phy-2325080; in part by the US Department of Energy Office of Advanced Scientific Computing Research, Accelerated Research for Quantum Computing Program under Award Number DE-SC0025509; and in part by the NSF Quantum Leap Challenge Institute for Hybrid Quantum Architectures and Networks (NSF Award 2016136), in part by the NSF National Virtual Quantum Laboratory program, in part based upon work supported by the U.S. Department of Energy, Office of Science, National Quantum Information Science Research Centers, and in part by the Army Research Office under Grant Number W911NF-23-1-0077. The views and conclusions contained in this document are those of the authors and should not be interpreted as representing the official policies, either expressed or implied, of the U.S. Government. The U.S. Government is authorized to reproduce and distribute reprints for Government purposes notwithstanding any copyright notation herein.

This research used resources of the National Energy Research Scientific Computing Center, a DOE Office of Science User Facility using NERSC award NERSC DDR-ERCAP0032212.

FTC is the Chief Scientist for Quantum Software at Infleqtion.

\appendix
\counterwithin*{equation}{section}
\renewcommand{\theequation}{\Alph{section}\arabic{equation}}\section{Coherent Error Model and Pauli Twirling}
\label{app:coherent-error-model}

\subsection{Coherent CZ Errors}
\label{app:coherent-cz-errors}
Gross et al.~\cite{gross2024characterizing} used a five-parameter model to characterize CZ gates on a Google Sycamore processor. In our simulations, we write an imperfect CZ gate as $\widetilde U_j=\mathrm{CZ}\,e^{-iH_j(\delta\boldsymbol{\theta}_j)}$, where $H_j(\boldsymbol u)=\sum_{b=1}^{5}u_b h_b$. For the five control coordinates $(\phi_j,x_j,y_j,\zeta_j,\gamma_j)$, we use the following generators:
\begin{equation}
\begin{aligned}
h_1 &= |11\rangle\langle11|, & h_2 &= \tfrac{XX+YY}{2}, & h_3 &= \tfrac{XY-YX}{2}, \\
&& h_4 &= \tfrac{ZI-IZ}{2}, & h_5 &= \tfrac{ZI+IZ}{2}.
\end{aligned}
\label{eq:cz-generators}
\end{equation}
Here $XY$ denotes $X\otimes Y$, and likewise for the other Pauli products. Each generator commutes with the CZ operator, so repeating the imperfect gate $\ell$ times produces an effective control offset $\ell\,\delta\boldsymbol{\theta}_j$. Applying $\pm(\lambda/\ell)\boldsymbol v_j$ to each copy keeps the total perturbation fixed at $\pm\lambda\boldsymbol v_j$. For odd $\ell$, the amplified unitary is $\widetilde U_{j,\ell}^{\pm}=\mathrm{CZ}\,e^{-iH_j(\ell\,\delta\boldsymbol{\theta}_j\pm\lambda\boldsymbol v_j)}$, where only the probed coordinate of $\boldsymbol v_j$ is nonzero.

\subsection{Pauli Twirling of an Amplified Block}
\label{app:block-twirling}
For either probe polarity, let $\widetilde U$ denote the amplified unitary $\widetilde U_{j,\ell}^{\pm}$ for $\ell$ copies of CZ gate $j$. Its error relative to the ideal CZ gate is $E=\mathrm{CZ}^{\dagger}\widetilde U$. Pauli twirling removes the cross terms between distinct Pauli operators in the error channel~\cite{emerson2007symmetrized,geller2013efficient}, leaving
\begin{equation}
\mathcal T(\rho)=\sum_{P\in\mathcal P_2}p_P P\rho P,\qquad
p_P=\left|\frac{\operatorname{Tr}(PE)}{4}\right|^2,
\label{eq:twirled-block-channel}
\end{equation}
where $\mathcal P_2=\{I,X,Y,Z\}^{\otimes 2}$. Because we compute $E$ for the amplified block, these Pauli probabilities account for coherent accumulation.

\section{Amplified Gradient Estimation and Variance}
\label{app:gradient-estimation}

\subsection{Normalization of the Amplified Response}
\label{app:amplification-normalization}

Consider one control parameter with offset $\delta\theta$ from its calibrated value. As shown in Appendix~\ref{app:coherent-error-model}, repeating the gate $\ell$ times increases the accumulated offset to $\ell\delta\theta$. Applying a perturbation of $\pm\lambda/\ell$ to each copy gives accumulated offsets of $\ell\delta\theta\pm\lambda$.

For small offsets, the local detection rate is approximately quadratic~\cite{sivak2026reinforcement,gong2026provably}. Its quadratic contribution is $\frac{\kappa}{2}\delta\theta^2$, where $\kappa$ describes the curvature, and its gradient is $\kappa\delta\theta$. We assume amplification does not substantially change the curvature; any change would bias the gradient estimate. When both probe settings remain in this quadratic regime,
\begin{equation}
\begin{aligned}
c_+ - c_- \approx \frac{\kappa}{2}\left[(\ell\delta\theta+\lambda)^2-(\ell\delta\theta-\lambda)^2\right] =2\kappa\lambda\ell\delta\theta.
\end{aligned}
\label{eq:amplified-quadratic-difference}
\end{equation}
Dividing this difference between the two detection rates by $2\lambda$ estimates the amplified gradient at $\ell \delta \theta$. To estimate the desired unamplified gradient, we divide by an additional factor of $\ell$.

For multiple controls, the same normalization applies to the gradient with respect to each control parameter. The controller probes each parameter in turn, measuring gates in the same probing group simultaneously. Collecting these estimates across all parameters and probing groups gives the full control gradient.

\subsection{Sampling Variance}
\label{app:sampling-variance}

For fixed probe signs $\boldsymbol v$ and amplification depth $\ell$, let $\sigma_\pm^2(\ell)$ denote the variance of the local contribution measured in one memory shot at the corresponding probe polarity. With $B$ independent memory shots per polarity, the sampling variance of the normalized gradient estimate is
\begin{equation}
\operatorname{Var}\left(\hat g(\ell)\mid\boldsymbol v\right)=\frac{\sigma_+^2(\ell)+\sigma_-^2(\ell)}{4B\lambda^2\ell^2}.
\label{eq:conditional-gradient-variance}
\end{equation}
If the per-shot variances $\sigma_\pm^2(\ell)$ are approximately constant, the gradient variance scales as $1/(B\lambda^2\ell^2)$, as in Eq.~\eqref{eq:gradient-variance-scaling}.

\subsection{Gradient Bias}
\label{app:gradient-accuracy}

Each pair of probe batches gives one gradient estimate $\hat g$ for a given gate and control parameter. Its bias is $b=\mathbb{E}[\hat g]-g$, where $g$ is the reference gradient and $\mathbb{E}[\hat g]$ is the expected estimate at fixed controls and probe settings. Let $\langle\cdot\rangle$ denote the average over all gates and their control parameters. The RMS bias is $b_{\mathrm{rms}}=\sqrt{\langle b^2\rangle}$, and the RMS reference gradient is $g_{\mathrm{rms}}=\sqrt{\langle g^2\rangle}$.

We compute $\mathbb{E}[\hat g]$ directly from the circuit model's detector probabilities at the two probe settings. We estimate the sampling variance by averaging $(\hat g-\mathbb{E}[\hat g])^2$ over 128 simulated gradient estimates per setting. For the SNR in Fig.~\ref{fig:snr-scaling}, we combine the sampling variance and squared bias to obtain the RMS error
\[
e_{\mathrm{rms}}
=\sqrt{\bigl\langle\mathbb{E}[(\hat g-g)^2]\bigr\rangle}
=\sqrt{\bigl\langle\operatorname{Var}(\hat g)+b^2\bigr\rangle}.
\]

\begin{figure}[t]
    \centering
    \includegraphics[width=0.8\columnwidth]{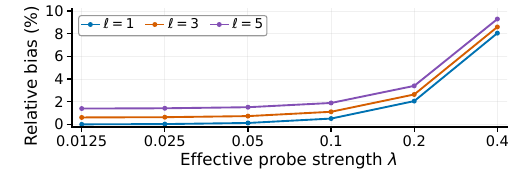}
    \caption{Relative RMS gradient bias, $(b_{\mathrm{rms}}/g_{\mathrm{rms}})\times 100\%$, versus effective probe strength.}
    \label{fig:gradient-bias}
\end{figure}

Figure~\ref{fig:gradient-bias} shows the relative RMS bias, $b_{\mathrm{rms}}/g_{\mathrm{rms}}$, as a function of probe strength. At all three amplification depths, it is below $2\%$ at $\lambda=0.1$ and exceeds $8\%$ at $\lambda=0.4$.

\section{Spatial Probing Groups}
\label{app:spatial-probing}

For gate $j$, let $\mathcal F_j$ be the set of affected detector coordinates and $\mathcal Q_j$ the set of data qubits that can carry a fault propagated from that gate. Two gates $j$ and $k$ in the same temporal layer can share a probing group if their detector footprints are disjoint and their data qubit footprints are separated by more than the dilution radius $R$. These conditions are
\begin{equation}
\mathcal F_j\cap\mathcal F_k=\varnothing,\qquad
\min_{q\in\mathcal Q_j, q'\in\mathcal Q_k}
\left\|\boldsymbol x_q-\boldsymbol x_{q'}\right\|_\infty>R,
\label{eq:spatial-group-constraints}
\end{equation}
where $\boldsymbol x_q$ is the lattice coordinate of qubit $q$, and the infinity norm gives the Chebyshev distance.

Within each layer, we connect two gates by an edge if they violate either condition in Eq.~\eqref{eq:spatial-group-constraints}. We color the resulting graph using DSATUR~\cite{brelaz1979new}. Gates of the same color form one probing group. The total number of groups across all layers is $W$.

\section{Batch Size Sensitivity}
\label{app:batch-size-sensitivity}

\begin{figure}[!t]
    \centering
    \includegraphics[width=0.8\columnwidth]{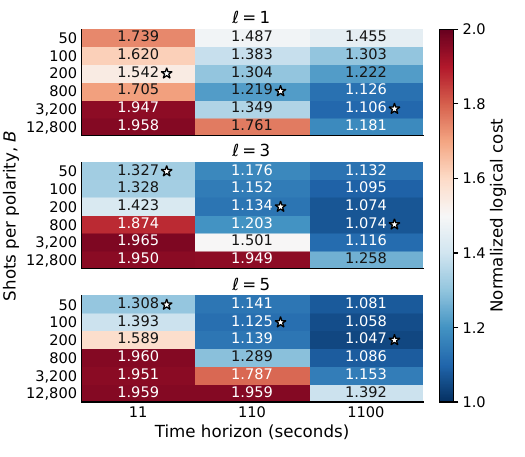}
    \caption{Normalized logical cost by batch size and execution window. For each batch size, we chose when to stop probing to minimize the logical cost over the window. Stars mark the batch size with the lowest cost for each time horizon.}
    \label{fig:validation-heatmap}
\end{figure}

For each execution window $T$ in Fig.~\ref{fig:fixed-window}, we examine how the logical cost varies with batch size $B$. At amplification depth $\ell$, the normalized logical cost is
\begin{equation}
    \widetilde K_\ell(T,B)
    =
    \frac{K_\ell(T,B)}{K_{\mathrm{ref}}(T)}.
    \label{eq:normalized-total-cost}
\end{equation}
Here $K_\ell(T,B)$ is the expected number of logical failures during probing and subsequent operation. For the reference cost $K_{\mathrm{ref}}(T)$, the calibrated circuit runs without probing for the full window, so only background noise contributes to logical failures. Fig.~\ref{fig:validation-heatmap} shows the normalized cost across batch sizes for each amplification depth.

\section{Detection Rates and Adaptive Probing}
\label{app:rho}

One possible adaptive probing rule compares the current detection-event rate with its average since the last probing episode began. Let $N_{\mathrm{det}}$ be the number of events recorded in that time and $T$ the elapsed time, including probing. During ordinary operation, the controller estimates the current event rate $r$ and could resume probing when $r\geq N_{\mathrm{det}}/T$. To apply this rule to a gate or probing group, we can compute $r$ and $N_{\mathrm{det}}$ using only detectors sensitive to that gate or group and measure $T$ from the start of its most recent probing episode. However, the time needed to estimate the current detection rate would limit how quickly the controller can respond to drift.

\bibliographystyle{ACM-Reference-Format}
\bibliography{refs}

\end{document}